\documentclass{revtex4}
\usepackage{graphicx}
\usepackage{fancyhdr}
\usepackage{amsmath}
\usepackage{color}

\begin{document}

\title{
Radiative corrections to the Higgs boson couplings in the Higgs triplet 
model\footnote{This proceedings is based on Ref.~\cite{AKKY}.}} 

%

\author{Mariko Kikuchi}
\affiliation{Department of Physics, University of Toyama, 3190 Gofuku, Toyama 930-8555, JAPAN}

\begin{abstract}
We calculate Higgs coupling constants at one-loop level in the Higgs triplet model (HTM) to compare to future collider experiments.
We evaluate the decay rate of the standard model (SM)-like Higgs boson ($h$) into diphoton.
Renormalized Higgs couplings with the weak gauge bosons $hVV$ ($V=W$\  and\ $Z$) and the trilinear coupling $hhh$ are also calculated at the one-loop level in the on-shell scheme.
The event rate of the $pp\rightarrow h \rightarrow \gamma\gamma$\ channel in the HTM to the one in the SM can cover the value from the recent LHC data. 
We find that in the allowed parameter region by the current data, deviations in the one-loop corrected $hVV$ and $hhh$ vertices can be about $-1\%$   and $+50\%$, respectively.
Magnitudes of these deviations can be enough significant to compare with the precision future data at the International Linear Collider. 

\end{abstract}

\maketitle

\thispagestyle{fancy}


\section{Introduction}
The Higgs boson whose the mass is about 126 GeV has been discovered via the $h\rightarrow \gamma\gamma, h\rightarrow ZZ^*\rightarrow 4\ell$ and $h\rightarrow WW^* \rightarrow \ell\nu \ell\nu$ channels with 5.9 $\sigma$ at ATLAS \cite{Higgs_ATLAS} and with 5.0 $\sigma$ at the CMS \cite{Higgs_CMS}.
It looks like the standard model (SM)-like Higgs boson.
However, it is not necessary that the Higgs boson is that of the SM. 
The SM-like Higgs boson can also be predicted in various extended Higgs sectors; e.g., 
the Higgs sector with additional SU(2) singlets, doublets and/or triplets.   
Such non-minimal Higgs sectors are introduced in various scenarios of
new physics beyond the SM which are motivated to solve the problems such as tiny neutrino masses, dark matter and/or 
 baryon asymmetry of the Universe.
Changing the viewpoints, if the true Higgs sector is determined by experiments, the new physics can also be determined.
Therefore, we can explore the new physics via the Higgs physics.

Discovery of new non-SM particles, such as charged Higgs bosons, CP-odd Higgs bosons and super particles, directly means new physics.
Currently the discovered new particle at LHC is only the SM-like Higgs boson.
Thus, studying this SM-like Higgs boson in detail is very important in order to determine the Higgs sector.  
We expect that the deviations in coupling constants of the SM-like Higgs boson from the SM predictions are detected 
at the LHC or at the future precision 
collider experiments such as the LHC at the integrated luminosity of 3000 fb$^{-1}$ and the International Linear Collider (ILC). 
Therefore, we can discriminate models of new physics by comparing accurate predictions on the coupling constants associated with 
the SM-like Higgs boson with the future precision measurements, even if additional new particles will be directly unfound.  

In this talk, we focus on the minimal Higgs triplet model (HTM).
This model can generate tiny neutrino masses via the so-called type-II seesaw mechanism \cite{typeII}.
One of the important feature in this model is that the electroweak rho parameter at the tree level ($\rho_{\textrm{tree}}$) deviates from unity due to the nonzero vacuum expectation value (VEV) of the triplet field $v_{\Delta}$.
First, we define a full-set of on-shell renormalization conditions.
The renormalization scheme in the HTM is different from the one in the SM because of the relation $\rho_{\textrm{tree}}=1$ does not hold.
Since an additional input parameter is required in electroweak sector, we must define one extra renormalization condition to determine the counter-term which corresponds to the additional input parameter.
Then, we calculate Higgs coupling constants at one-loop level; e.g., $h\gamma\gamma,\ hZZ,\ hWW$ and the Higgs triple coupling $hhh$.
We evaluate deviations in these coupling constants from the predictions in the SM under the allowed parameter regions by the electroweak precision data and bounds from perturbative unitarity and vacuum stability.
We then discuss the possibility to test the HTM by comparing these calculations with future precision data at collider experiments.

\section{HIGGS TRIPLET MODEL}
The scalar sector of the HTM is composed of the isospin doublet field $\Phi$ with 
hypercharge $Y=1/2$ and the triplet field $\Delta$ with $Y=1$. 
The detail of Lagrangian is given in Ref.~\cite{typeII}.
The electroweak rho parameter $\rho$ is given as the following form, 
\begin{align}
\rho \equiv \frac{m_W^2}{m_Z^2\cos^2\theta_W}=\frac{1+\frac{2v_\Delta^2}{v_\phi^2}}{1+\frac{4v_\Delta^2}{v_\phi^2}}, \label{rho_triplet}
\end{align}
where $v_{\phi}$ and $v_{\Delta}$ are the VEVs of the doublet Higgs field and the triplet Higgs field, respectively, which satisfy the relation $v^2 \equiv v_{\phi}^2 + 2v_{\Delta}^2 \simeq (246\ \textrm{GeV})^2$.
Namely, $\rho$ deviates from unity at the tree level. 
The experimental value of the rho parameter is quite close to unity; i.e., $\rho^{\text{exp}}=1.0008^{+0.0017}_{-0.0007}$~\cite{PDG}, 
so that 
$v_\Delta$ has to be less than about 8 GeV by using the tree level formula in Eq.~(\ref{rho_triplet}).

This model has the new interaction for neutrinos \cite{typeII}.
It is the one between the triplet field and lefthand neutrinos.
First, two lefthand neutrinos couple to $\Delta$, then $\Delta$ carries the lepton number of $-2$.
When it couples to two $\Phi$, the lepton number is broken at the vertex.
Neutrino masses of the Majorana type are produced by these interactions.

The most general form of the Higgs potential under the gauge symmetry is given by 
\begin{align}
V(\Phi,\Delta)&=m^2\Phi^\dagger\Phi+M^2\text{Tr}(\Delta^\dagger\Delta)+\left[\mu \Phi^Ti\tau_2\Delta^\dagger \Phi+\text{h.c.}\right]\notag\\
&+\lambda_1(\Phi^\dagger\Phi)^2+\lambda_2\left[\text{Tr}(\Delta^\dagger\Delta)\right]^2+\lambda_3\text{Tr}[(\Delta^\dagger\Delta)^2]
+\lambda_4(\Phi^\dagger\Phi)\text{Tr}(\Delta^\dagger\Delta)+\lambda_5\Phi^\dagger\Delta\Delta^\dagger\Phi, \label{pot_htm}
\end{align}
where $m$ and $M$ are dimension full real parameters, $\mu$ is the dimension full complex parameter 
which violates the lepton number, and 
$\lambda_1$-$\lambda_5$ are the coupling constants. 
We here take $\mu$ to be real. 
There are seven physical mass eigenstates $H^{\pm\pm}$, $H^\pm$, $A$, $H$ and $h$ 
in addition to the three NG bosons $G^\pm$ and $G^0$ which are absorbed by the longitudinal components 
of the $W$ boson and the $Z$ boson.

When $v_\Delta$ is much less than $v_\phi$, which is required by the rho parameter data, 
there appear relationships\cite{AKKY,AKKY1} among the masses of the triplet-like Higgs bosons by neglecting $\mathcal{O}(v_\Delta^2/v_\phi^2)$ terms as 
\begin{align}
m_{H^{++}}^2-m_{H^{+}}^2&=m_{H^{+}}^2-m_{A}^2~~\left(=-\frac{\lambda_5}{4}v^2\right), ~\label{eq:mass_relation1}\\
m_A^2&=m_{H}^2~~(=M_\Delta^2). \label{eq:mass_relation2}
\end{align}
Notice that mass hierarchy among the triplet-like Higgs bosons depends on the sign of $\lambda_5$.
If $\lambda_5$ is positive(negative), $H^{++}$($A$ and $H$) is the lightest of all the triplet-like Higgs bosons; i.e., $m_A\, >\, m_{H^+}\, >m_{H^{++}}$($m_{H^{++}}\, >\, m_{H^+}\, >m_{A}$)\cite{lambda5,AKKY,KY,AKKY1}.
We call the former case (latter case) as Case~I (Case~II). 
We define $\Delta m$ as the mass difference between the singly charged Higgs boson and the lightest triplet-like Higgs boson; i.e., $\Delta m\equiv m_{H^+}-m_{\textrm{lightest}}$.

\section{RENORMALIZATION CALCULATION}
We here define on-shell renormalization conditions in this model.
First, we discuss the renormalization of the electroweak sector to calculate the renormalized $W$ boson mass, 
which can be used to constrain parameters such as the triplet-like Higgs boson masses. 
Second, we discuss the renormalization of parameters in the Higgs potential.

\subsection{Electroweak parameters}
There are five electroweak parameters, $m_W,\ m_Z,\ \sin\theta_W,\ G_F$ and $\alpha_{\textrm{em}}$, in the model with $\rho_{\textrm{tree}}=1$.
They are described by three independent input parameters.
For instance, when we chose $m_W,\ m_Z$ and $\alpha_{\textrm{em}}$ as input parameters, all the other parameters are written in these input parameters~\cite{Hollik-SM}.
Each counter-terms can be determined by imposing renormalization conditions.
Counter-terms of $m_W$ and $m_Z$ can be determined by on-shell conditions for two point functions, and the one of $\alpha_{\textrm{em}}$ is determined by on-shell conditions for the $ee\gamma$ vertex~\cite{Hollik-SM,KY,AKKY}.

On the other hand, in the HTM, four input parameters are required to be fixed because the relation $\rho_{\textrm{tree}}=1$ does not hold.
Therefore, we need an additional input parameter.
Here, three of four input parameters are chosen from the electroweak precision observables; i.e., $m_W,\ m_Z$ and $\alpha_{\textrm{em}}$ as in the SM.
We chose $\theta_W$ as the other one, which $\theta_W$ is related the mixing angle $\beta'$ among CP-odd scalar bosons by
\begin{align}
\cos^2\theta_W = \frac{2m_W^2}{m_Z^2(1+\cos^2\beta')}. \label{swsq_2}
\end{align}
We determine the counter-term of $\theta_W$ by using this relation from putting the condition on $\beta'$.
This is the difference in the renormalization scheme between the model with $\rho_{\textrm{tree}}=1$ and the HTM.
Renormalized $W$ boson mass is calculated by these renormalization conditions in Ref.~\cite{AKKY}.
We find that the mass difference $\Delta m$ is constrained by the LEP/SLC electroweak precision data~\cite{PDG} as $0<\Delta m \lesssim 50$ GeV ($0<\Delta m \lesssim 30\text{ GeV}$) for $v_\Delta \lesssim 1$ GeV, 
$40\text{ GeV}\lesssim \Delta m \lesssim 60$ GeV ($30\text{ GeV}\lesssim \Delta \lesssim 50$ GeV) for $v_\Delta=5$ GeV 
and $85\text{ GeV}\lesssim \Delta m\lesssim 100$ GeV ($70\text{ GeV}\lesssim \Delta m\lesssim 85$ GeV) 
for $v_\Delta=10$ GeV.

\subsection{Higgs potential}
There are nine parameters in the Higgs potential ($v,\ \alpha,\ \beta,\ \beta',\ m_{H^{++}},\ m_{H^+},\ m_A,\ m_H,\ m_h$, where $\alpha\ (\beta)$ is the mixing angle among  CP-even (charged) scalar bosons).
We determine the counter-term of $v$ by the renormalization in the electroweak parameters.
$\beta$ is determined through the relation with $\beta'$.
Other counter-terms can be determined by the on-shell conditions in the Higgs potential renormalization~\cite{AKKY1}. 
The detail of this renormalization is described in the Ref.~\cite{AKKY}.

\section{HIGGS COUPLINGS AT THE ONE- LOOP LEVEL}
In this section, we discuss the SM-like Higgs boson ($h)$ couplings with the gauge bosons 
($\gamma\gamma$, $W^+W^-$ and $ZZ$) and the Higgs selfcoupling $hhh$ at the one-loop level in the favored 
parameter regions by the unitarity bound, the vacuum stability bound and the 
measured $W$ boson mass discussed in previous sections.
The mass difference $\Delta m$ is constrained from the perturbative unitarity and the vacuum stability because $\Delta m$ depends on $\lambda_4$ and $\lambda_5$.
The condition for the vacuum stability bound has been derived in Ref.~\cite{Arhrib}, where 
we require that 
the Higgs potential is bounded from below in any directions.
The unitarity bound has been discussed in Ref.~\cite{Aoki-Kanemura} in the Gerogi-Machacek model~\cite{GM} 
which contains the HTM. The unitarity bound in the HTM has also been derived in Ref.~\cite{Arhrib}.

\begin{figure}
\begin{center}
\includegraphics[width=70mm]{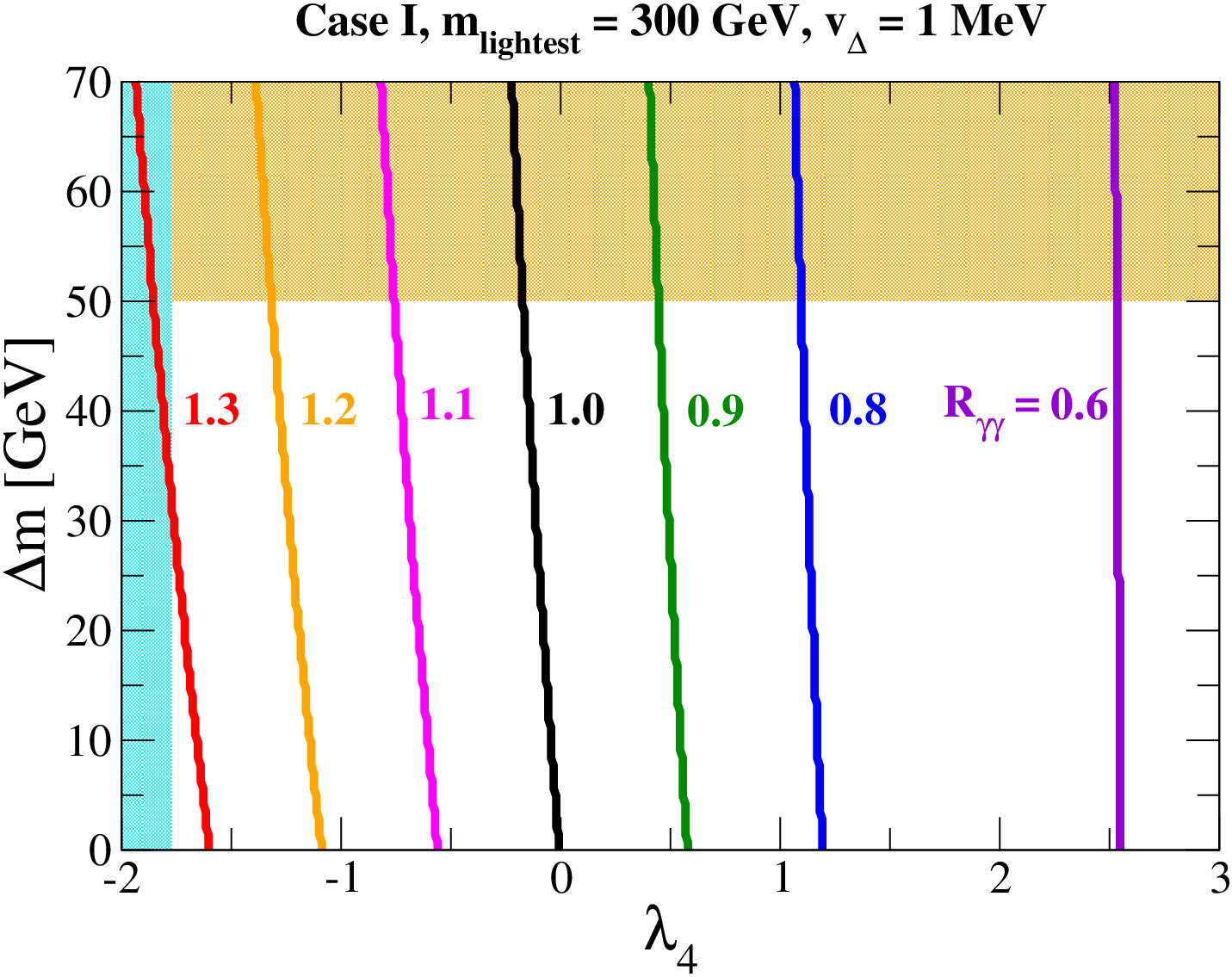}\hspace{3mm}
\includegraphics[width=70mm]{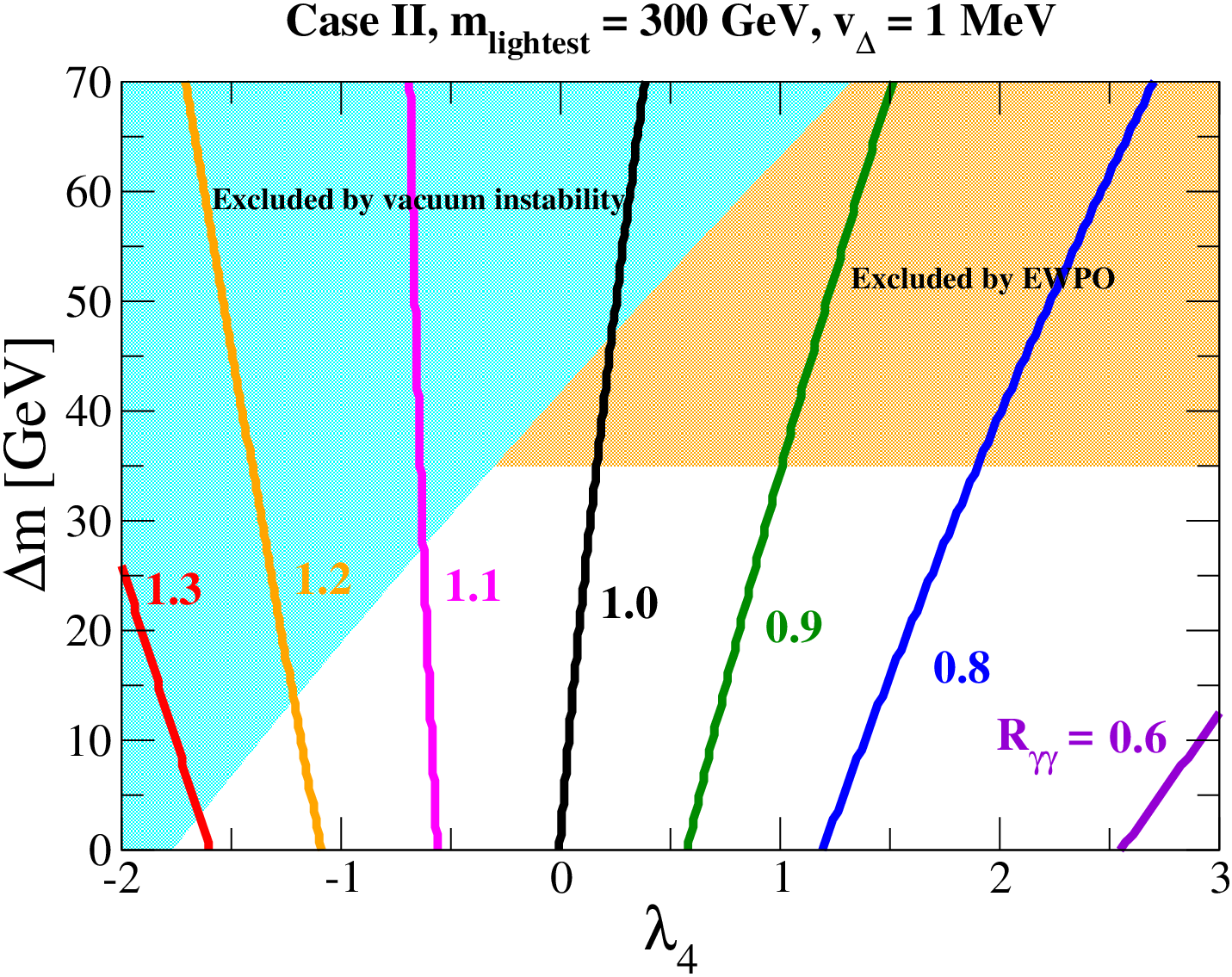}
\caption{Contour plots of $R_{\gamma\gamma}$ for $v_\Delta=1$ MeV and $m_{\text{lightest}}=300$ GeV 
in the $\lambda_4$-$\Delta m$ plane. 
The left panel (right panel) shows the result in Case~I (Case~II).  
The blue and orange shaded regions are excluded by the vacuum stability bound and the measured $m_W$ data, respectively. 
}
\label{FIG:hgg}
\end{center}
\end{figure}

First, we discuss the decay of the diphoton channel: $h\to\gamma\gamma$~\cite{AKKY, KY, h_to_gg}, which is important in the Higgs boson search at the LHC. 
The current experimental value of the signal strength for the Higgs to diphoton mode is $1.6\pm0.3$ at the ATLAS~\cite{ATLAS_new} and
$0.8\pm0.3$ at the CMS~\cite{CMS_new}. 
We can directly detect new charged particles on the loop via $h\rightarrow \gamma\gamma$ process because this process is the one-loop process. 
In the HTM, the doubly-charged Higgs boson $H^{\pm\pm}$ and the singly-charged Higgs boson $H^\pm$ 
can contribute to the diphoton decay.  
In particular, 
the contribution from the $H^{\pm\pm}$ loop to the $h\to \gamma\gamma$
is quite important compared to that from $H^{\pm}$, because $H^{\pm\pm}$ 
contribution is roughly 4 times larger than that from the $H^{\pm}$ contribution at the amplitude level. 
Then, we evaluate the ratio of the event rate for $h\rightarrow \gamma\gamma$ in the HTM to that in the SM, taking into account the constraint from the perturbative unitarity, the vacuum stability and the electroweak precision data.
We define it as the following:
\begin{align}
R_{\gamma\gamma}\equiv 
\frac{\sigma(gg\to h)_{\text{HTM}}\times 
\text{BR}(h\to \gamma\gamma)_{\text{HTM}}}{\sigma(gg\to h)_{\text{SM}}\times \text{BR}(h\to \gamma\gamma)_{\text{SM}}}, 
\end{align}
where $\sigma(gg\to h)_{\text{model}}$ is the cross section of the gluon fusion process, and 
$\text{BR}(h\to \gamma\gamma)_{\text{model}}$ is the branching fraction of the $h\to \gamma\gamma$ mode in a model. 
In fact, the ratio of the cross section $\sigma(gg\to h)_{\text{HTM}}/\sigma(gg\to h)_{\text{SM}}$ can be 
replaced by the factor $\cos^2\alpha/\cos^2\beta$. 
In Fig.~\ref{FIG:hgg}, we show the contour plots of $R_{\gamma\gamma}$ for $v_\Delta=1$ MeV and $m_{\text{lightest}}=300$ GeV
on the $\lambda_4$-$\Delta m$ plane. 
The left panel (right panel) shows the result in Case~I (Case~II).  
The blue and orange shaded regions are those excluded by the vacuum stability bound (assuming $\lambda_{2,\ 3}=3$) 
and the measured $m_W$ data, respectively. 
In this model, $R_{\gamma\gamma}$ is very  sensitive to $\lambda_4$ because SM-like  Higgs boson couplings with charged Higgs bosons are composed of $\lambda_4$~\cite{AKKY, h_to_gg}.
We note that the dependence for $\Delta m$ of $R_{\gamma\gamma}$ in Case I is small because $m_{H^{++}}$ is fixed.
On the other hand, the result in Case II slightly depends on $\Delta m$ due to the larger values of $m_{H^{++}}$ which affects $R_{\gamma\gamma}$ via $\Delta m$. 
Under the constraint of the vacuum stability and the electroweak precision observable $m_W$, larger 
$\Delta m$ can be allowed in Case I than in Case II. 
We find that predicted values of $R_{\gamma\gamma}$ are about 1.3 (about 0.6) in this case when $\lambda_4$ is about $-1.7$ (about $3$) in both Case~I and Case~II. 
The data at the ATLAS is rather different from those at the CMS.
When we take into account the CMS data, the parameter region $\lambda_4 \gtrsim -0.5$ is favored.

Next, we calculate the Higgs coupling constants at the one-loop level by the renormalization which we discuss at the previous section.
Then, we define following quantity to study deviations for $hVV$ and $hhh$ coupling from the SM predictions: 
\begin{align}
\Delta g_{hVV} \equiv \frac{\text{Re}M_1^{hVV}-\text{Re}M_1^{hVV}(\text{SM})}{\text{Re}M_1^{hVV}(\text{SM})},  \label{hgVV}
\end{align}
where $M_1^{hVV}$ is the form factor of the $hVV$ coupling in the HTM, which is proportional to the Minkowshi's metric tensor $g^{\mu\nu}$.
$M_1^{hVV}(\text{SM})$ is the corresponding prediction in the SM.
We fix values of momenta such as $p_1=m_V,\ p_2=m_h-m_V$ and $q=m_h$, where $p_1$ and $p_2$ are external incoming momenta and $q$ is the outgoing momentum.
\begin{align}
\Delta \Gamma_{hhh}\equiv \frac{\text{Re}\Gamma_{hhh}-\text{Re}\Gamma_{hhh}^{\text{SM}}}{\text{Re}\Gamma_{hhh}^{\text{SM}}}, \label{delhhh}
\end{align}
where $\Gamma_{hhh}$ is the form factor of the $hhh$ coupling in the HTM, and
$\Gamma_{hhh}^{\textrm{SM}}$ is the corresponding prediction in the SM.
We fix values of momenta such as $p_1=m_h,\ p_2=m_h$ and $q=2m_h$.

The deviation for the $hWW$ coupling $\Delta g_{hWW}$ is predicted to be at most a few percent in the allowed parameter regions by the vacuum stability and by the measured $W$ boson mass in Case I and Case II.
Even if we take into account the LHC data of the signal strength for the diphoton mode, $\Delta g_{hWW}$ can be about 1\%.
The results for deviations for $hZZ$ coupling $\Delta g_{hZZ}$ are very similar to these for $\Delta g_{hWW}$.
Deviations in $hVV$ are expected to be measured at the ILC with a center of mass energy 
to be 500 GeV and integrated luminosity being 500 fb$^{-1}$~\cite{Peskin}.

\begin{figure}
\begin{center}
\includegraphics[width=70mm]{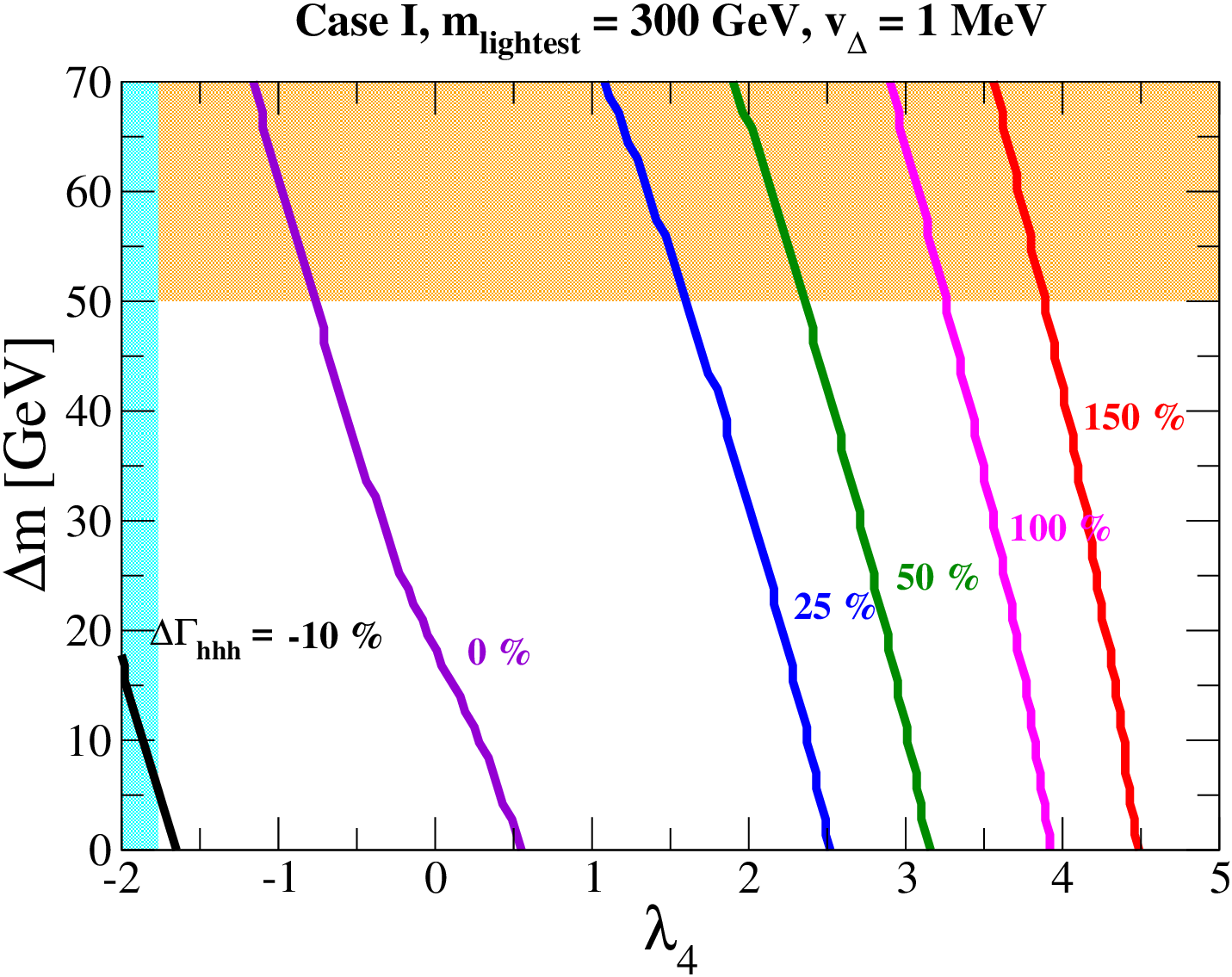}\hspace{3mm}
\includegraphics[width=70mm]{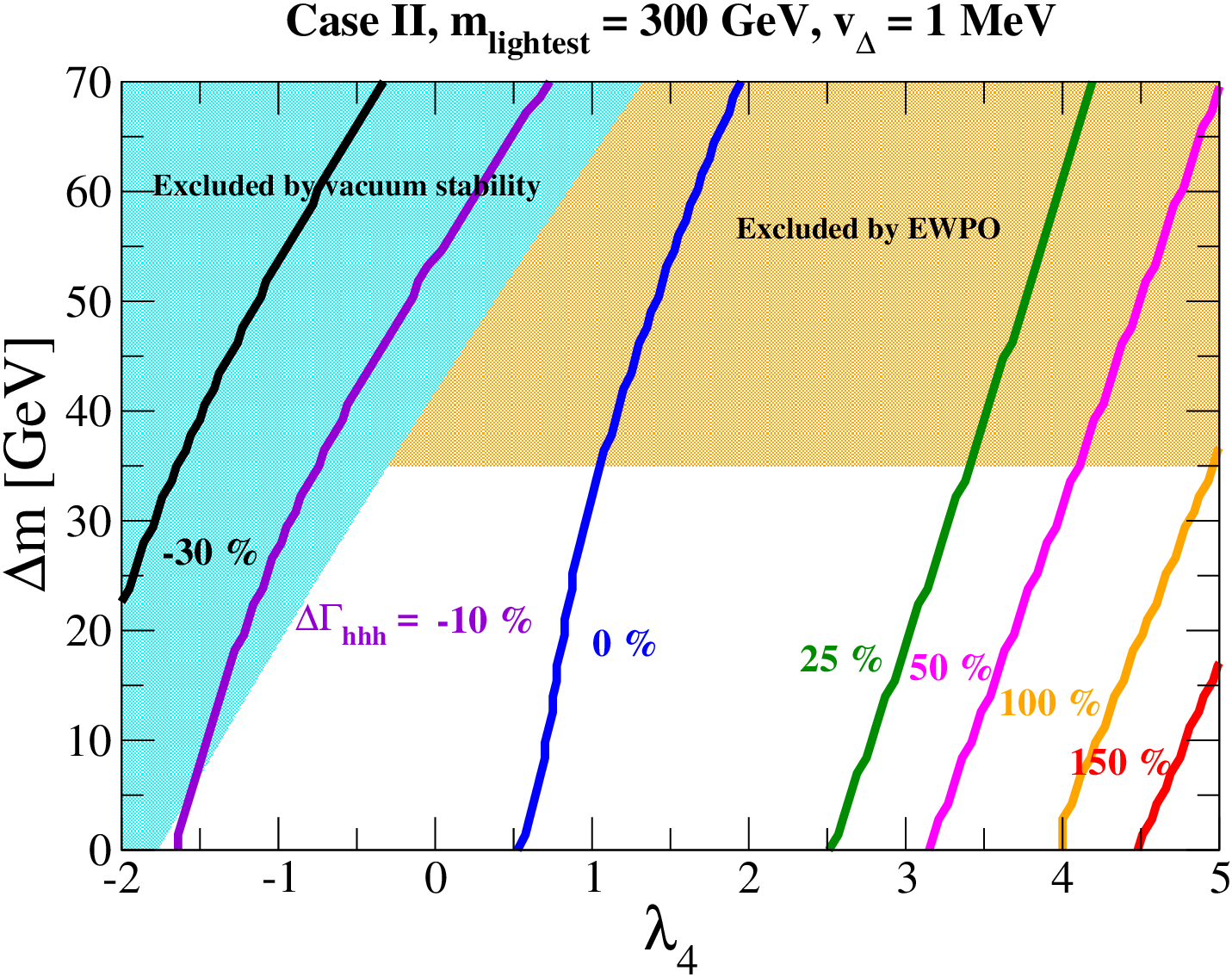}
\caption{Contour plots of
$\Delta\Gamma_{hhh}$ defined 
in Eq.~(\ref{delhhh}) for $m_{\text{lightest}}=300$ GeV and $v_\Delta=1$ MeV. 
The left panel (right panel) shows the result in Case~I (Case~II).  
The blue and orange shaded regions are excluded by the vacuum stability bound and the measured $m_W$ data, respectively. }
\label{FIG:hhh}
\end{center}
\end{figure}

In Fig.~\ref{FIG:hhh}, 
the contour plot for 
the deviation of $hhh$ coupling  $\Delta\Gamma_{hhh}$ defined 
in Eq.~(\ref{delhhh}) is shown for $m_{\text{lightest}}=300$ GeV and $v_\Delta=1$ MeV
in the $\lambda_4$-$\Delta m$ plane. 
The left (right) plot shows the result in Case~I (Case~II).  
The blue and orange shaded regions are excluded by the vacuum stability bound and the measured $m_W$ data, respectively. 
In both cases, positive (negative) values of $\Delta \Gamma_{hhh}$ are predicted in the case with a positive (negative) $\lambda_4$
whose magnitudes can be greater than about +150\% ($-10\%$) under the constraint from perturbative unitarity~\cite{Arhrib, Aoki-Kanemura}.
The large deviation in the $hhh$ coupling constant from the non-decoupling property of scalar bosons in the loop, as was well known in the case of two Higgs doublet model~\cite{THDM_reno}. 
Even if we take into account the LHC data of the signal strength for the diphoton mode, $\Delta \Gamma_{hhh}$ can be about $+50$\%. 
Such a deviation in $\Delta\Gamma_{hhh}$ is expected to be measured at the ILC with a center of mass energy 
to be 1 TeV and integrated luminosity being 2 ab$^{-1}$~\cite{Fujii}.

We find that Higgs couplings with the gauge bosons ($\gamma,\ W$ and $Z$) and the Higgs triple coupling with radiative corrections in the HTM may deviate from predictions in the SM. 
In particular, when $\lambda_3$ is close to $3$, deviations for these coupling constants are large enough to be measured these at the ILC.
Even if any of the coupling constants does not deviate much, we may obtain features in the HTM through the correlation among these coupling constants.
Namely, this model may be testable by comparing precise theoretical predictions on these coupling constants with precision measurements at future collider experiments, especially at the ILC.
   

 \section{CONCLUSIONS}
We have calculated some Higgs coupling constants at the one-loop level in the HTM in order to compare to the data at future collider experiments.
We have discussed the renormalization conditions in this model for one-loop calculations.  
We have computed the decay rate of the SM-like Higgs boson $h$ into diphoton. 
Renormalized Higgs couplings with the weak gauge bosons $hVV$ and 
the Higgs self-coupling $hhh$ have also been calculated at the one-loop level. 
Magnitudes of deviations in these quantities from predictions of the SM 
have been evaluated in the parameter regions where the unitarity and vacuum 
stability bounds are satisfied and the predicted $W$ boson mass is consistent with the data.
In the allowed region by the LHC data, deviations in the one-loop corrected $hVV$ and $hhh$ vertices can be about $-1\%$   and $+50\%$, respectively.
We can obtain features in the HTM by testing the pattern of deviations in coupling constants from the SM predictions.
The HTM may be distinguished from the other models, by comparing to measure these deviations in Higgs boson couplings accurately. 
These deviations in the Higgs boson couplings may be detected at future colliders such as 
the LHC with 3000 fb$^{-1}$ and at the ILC. 
\\\\

\noindent
$Acknowledgments$

I'm very grateful to Shinya Kanemura and Kei Yagyu for careful reading of the manuscript.



\bigskip 

\end{document}